\documentclass[twocolumn,secnumarabic,amssymb, nobibnotes, aps, prd, showpacs]{revtex4}
\usepackage{mathbbold}
\usepackage{graphicx}
\usepackage{dcolumn}
\usepackage{bm}
\usepackage{hyperref}
\begin{document}

\title{Gluon Condensate in Pion Superfluid beyond Mean Field Approximation}
\author{Yin Jiang}
\author{Pengfei Zhuang}
\affiliation{Physics Department, Tsinghua University, Beijing
100084, China}
\date{\today}

\begin{abstract}
We study gluon condensate in a pion superfluid, through calculating
the equation of state of the system in the Nambu--Jona-Lasinio
model. While in mean field approximation the growing pion condensate
leads to an increasing gluon condensate, meson fluctuations reduce
the gluon condensate and the broken scalar symmetry can be smoothly
restored at finite isospin density.
\end{abstract}


\maketitle

Quarks and gluons condense in the vacuum of Quantum Chromodynamics
(QCD). From lattice QCD calculations and effective QCD models in hot
medium, it is widely accepted that the quark condensate
$\langle\bar\psi\psi\rangle$ which is the order parameter of the
chiral symmetry restoration drops down at finite temperature. The
gluon condensate $\langle G^a_{\mu\nu}
G^{\mu\nu}_a\rangle$\cite{collins,fukuda} which describes the degree
of the scale symmetry breaking is, however, not so optimistic.

The gluon condensate at finite temperature is investigated in
instanton model\cite{gross}, renormalization group\cite{schaefer},
QCD sum rule\cite{mallik}, and effective QCD models at low
energy\cite{celenza,cohen,mishra,sollfrank,malheiro,baldo,agasian}.
While the results in these calculations are quantitatively
different, they show the same temperature trend of the gluon
condensate: It is almost invariable at low temperature and starts to
decrease rapidly around the critical temperature of QCD phase
transitions.

At finite isospin density, both the Lee-Huang-Yang model\cite{lhy}
for a dilute Boson gas and the Nambu--Jona-Lasinio (NJL)
model\cite{njl} show a surprising mean field result\cite{he}: In the
pion superfluid the gluon condensate drops down slightly only at
very low isospin density but goes up and even exceeds its vacuum
value when the density is high enough. This result is qualitatively
in agreement with the calculations for 2-color baryon matter and
3-color isospin matter\cite{metlitski,zhitnitsky}. A natural
question is if this conclusion is still true when we go beyond the
mean field. In this paper, we study the gluon condensate in pion
superfluid in the NJL model beyond mean field.

Neglecting the current quark mass $m$, the QCD Lagrangian is
invariable under the scale transformation
$\psi(x)\rightarrow\lambda^{3/2}\psi(\lambda x)$ for the quark field
and $A^\mu(x)\rightarrow\lambda A^\mu(\lambda x)$ for the gauge
field. At classical level, the trace of the corresponding Noether
current is $\partial^\mu J_\mu = T^\mu_\mu = m\overline\psi\psi$. At
quantum level, the running coupling constant $\alpha_s$ leads to a
so-called anomaly term, the trace of the ensemble average of $
T^\mu_\nu$ becomes exactly the trace of the energy-momentum tensor
of the system, and therefore the matter parts of the quark and gluon
condensates at finite temperature $T$ and chemical potential $\mu$
are related to the energy density $\epsilon$ and pressure $p$ of the
system\cite{drummond},
\begin{equation}
\label{thermal} \epsilon-3p =-{9\over 8}\langle {\alpha_s\over
\pi}G^a_{\mu\nu}G_a^{\mu\nu}\rangle_{T,\mu}+m\langle
\overline\psi\psi\rangle_{T,\mu}.
\end{equation}
This relation tells us that the QCD condensates are controlled by
the bulk properties of the system. Since it is difficult to directly
calculate the QCD thermodynamics in non-perturbative region, this
relation gives a way to qualitatively estimate the gluon condensate
in effective models at low energy where partons are not explicit
constituents, if the model can reasonably describe the QCD
thermodynamics. For instance, the gluon condensate has been
investigated in nuclear
matter\cite{cohen,mishra,malheiro,baldo,agasian} and in isospin
matter\cite{metlitski,he} with low-energy models. When we neglect
the current quark mass $m$, the gluon condensate decouples from the
quark condensate and is purely controlled by the thermodynamics of
the system. While the gluon condensate for an ideal gas with
$\epsilon-3p=0$ is medium independent, it will be significantly
changed for a strongly coupled system. From the lattice simulation
at finite temperature\cite{cheng}, the QCD system is a strongly
coupled matter around the phase transition temperature $T_c$ with
$\epsilon-3p \gg 0$. This is the reason why the gluon condensate
drops down dramatically around $T_c$.

The NJL model at quark level\cite{njlquark} has been successfully
used to study chiral symmetry restoration, color superconductivity
and pion superfluidity at moderate temperature and density. The
flavor SU(2) NJL model is defined through the Lagrangian density
\begin{equation}
\label{njl1} {\mathcal{L}} = \overline\psi\left(i\gamma^\mu
\partial_\mu-m+\mu\gamma_0\right)+G\left[\left(\overline\psi\psi\right)^2+\left(\overline\psi
i\gamma_5{\bf \tau}\psi\right)^2\right],
\end{equation}
where the quark chemical potential matrix
$\mu=diag(\mu_u,\mu_d)=diag(\mu_B/3+\mu_I/2,\mu_B/3-\mu_I/2)$ and
the Pauli matrices ${\bf \tau} = (\tau_1,\tau_2,\tau_3)$ are defined
in flavor space, $\mu_B$ and $\mu_I$ are baryon and isospin chemical
potentials, and $G$ is the four-fermion coupling constant. The NJL
thermodynamic potential can be separated into a mean field part and
a fluctuation part,
\begin{equation}
\label{omega} \Omega=\Omega_{MF}+\Omega_{FL}.
\end{equation}

The mean field part $\Omega_{MF}$ contains the mean field potential
and the contribution from the quasi-quarks\cite{he2},
\begin{eqnarray}
\label{mf} \Omega_{MF} &=&
G\left(\sigma^2+\pi^2\right)\nonumber\\
&-&3\int{d^3{\bf k}\over
(2\pi)^3}\left[E^+_-+E_-^--E^+_+-E_+^-\right]\nonumber\\
&+&2T\ln\left(1+e^{-E_-^+/T}\right)\left(1+e^{-E_-^-/T}\right)\nonumber\\
&+&2T\ln\left(1+e^{E_+^+/T}\right)\left(1+e^{E_+^-/T}\right),
\end{eqnarray}
where the chiral condensate $\sigma=\langle\overline\psi\psi\rangle$
and pion condensate $\pi=\sqrt 2\langle\overline\psi
i\gamma_5\tau_+\psi\rangle$ with $\tau_+=(\tau_1+i\tau_2)/\sqrt 2$
are determined by minimizing the potential,
\begin{equation}
\label{gaps}
{\partial\Omega_{MF}\over\partial\sigma}=0,\ \
{\partial\Omega_{MF}\over \partial\pi}=0,\ \
{\partial^2\Omega_{MF}\over \partial\sigma^2}>0,\ \
{\partial^2\Omega_{MF}\over \partial\pi^2}>0,
\end{equation}
and $E_\mp^\pm=E_k^\pm\mp\mu_B/3$ are the quasi-quark energies with
$E_k^\pm=\sqrt{\left(E_k\pm\mu_I/2\right)^2+4G^2\pi^2}$,
$E_k=\sqrt{{\bf k}^2+M_q^2}$ and dynamical quark mass
$M_q=m-2G\sigma$.

In the NJL model, the meson modes are regarded as quantum
fluctuations above the mean field. The two quark scattering via
meson exchange can be effectively expressed in terms of quark bubble
summation in random phase approximation\cite{njlquark}. In normal
phase without pion condensation, the bubble summation selects its
specific isospin channel by choosing at each stage the same proper
polarization, and the meson masses $M_m\ (m = \sigma, \pi_+, \pi_-,
\pi_0)$ which are determined by poles of the meson propagators,
$1-2G\Pi_{mm}(M_m,{\bf 0})= 0$, are related only to their own
polarization functions $\Pi_{mm}(q_0,{\bf q})$. In pole
approximation, the meson contribution to the thermodynamic potential
can be expressed as\cite{hufner}
\begin{eqnarray}
\label{normal} \Omega_{FL} &=& \sum_m \Omega_m,\\
\Omega_m &=& \int{d^3{\bf q}\over (2\pi)^3}\left[{E_m-\mu_m\over
2}+T\ln\left(1-e^{-{E_m-\mu_m\over T}}\right)\right]\nonumber
\end{eqnarray}
with meson energies $E_m=\sqrt{{\bf q}^2+M_m^2}$ and meson isospin
chemical potential $\mu_{\pi_\pm}=\pm\mu_I$ and
$\mu_{\pi_0}=\mu_\sigma=0$.

In the pion superfluid phase, the quark propagator contains
off-diagonal elements in flavor space, we must consider all possible
isospin channels in the bubble summation. In this case, all the
possible polarizations form a matrix $\Pi$ in the four-dimensional
meson isospin space with off-diagonal elements $\Pi_{mn}$. While
there is no mixing between $\pi_0$ and other mesons,
$\Pi_{\pi_0\sigma}=\Pi_{\pi_0\pi_+}=\Pi_{\pi_0\pi_-}=0$, the other
three mesons are coupled to each other. The explicit $T, \mu_B$ and
$\mu_I$ dependence of all polarization elements $\Pi_{mn}$ can be
found in Appendix B of \cite{he2}. When the system goes through the
phase transition line and enters the normal phase, all the
off-diagonal elements disappear automatically.

The masses of the eigen modes of the Hamiltonian ${\cal H}$ in the
pion superfluid are defined through the poles of the meson
propagator, det$\left(1-2G\Pi(M_\theta,{\bf 0})\right)=0$ which can
be separated into $1-2G\Pi_{\pi_0\pi_0}(M_{\pi_0},{\bf 0})=0$ for
$\theta=\overline\pi_0$ and det$\left(1-2G\Pi(M_\theta,{\bf
0})\right)=0$ in the three-dimensional isospin subspace for
$\theta=\overline\sigma, \overline\pi_+, \overline\pi_-$. Different
from the normal phase where the meson modes $\sigma, \pi_+, \pi_-,
\pi_0$ are eigen states of both the Hamiltonian ${\cal H}$ and the
isospin operator $\hat I_3=1/2\int d^3{\bf
x}\bar\psi\gamma_0\tau_3\psi$ of the system, only $\overline\pi_0$
is still the eigen state of $\hat I_3$ (we still label it $\pi_0$ in
the following), but $\overline\sigma, \overline\pi_+,
\overline\pi_-$ have no longer definite isospin quantum number. The
eigen states of $\hat I_3$ are only related to the diagonal elements
$\Pi_{mm}$ and their masses are defined by $1-2G\Pi_{mm}(M_m,{\bf
0})=0$.

After taking bubble summation and Matsubara frequency summation, the
fluctuation part of the thermodynamic potential can be generally
written as\cite{hufner}
\begin{eqnarray}
\label{superfluid} \Omega_{FL} &=& -\int{d^3{\bf q}\over
(2\pi)^3}\int_0^\infty {d\omega\over 2\pi i}\left[{\omega\over
2}+T\ln\left(1-e^{-\omega/T}\right)\right]\nonumber\\
&\times&{d\over d\omega}\ln{\text
{det}\left(1-2G\Pi(\omega+i\epsilon,{\bf q})\right)\over \text
{det}\left(1-2G\Pi(\omega-i\epsilon,{\bf q})\right)},
\end{eqnarray}
where the two polarization matrices are respectively defined in the
top and bottom complex meson energy plane. An often used
simplification to calculate $\Omega_{FL}$ is the pole approximation,
namely neglecting the scattering phase shifts and considering only
the contribution from the quasi particles, like (\ref{normal}) for
the normal phase. In this case, we have
$\Omega_{FL}=\sum_\theta\Omega_\theta$. To explicitly show the
isospin dependence, we further make a transformation\cite{hao} from
the basis $(\overline\sigma, \overline\pi_+, \overline\pi_-)$ to the
basis $(\sigma, \pi_+, \pi_-)$. The elemental states in the former
basis do not carry definite isospin quantum numbers, but the later
is constructed by the eigen states of the isospin operator $\hat
I_3$. Since the two spaces are both complete, such a transformation
will not lose any information. Taking into account the orthogonal
condition for the two spaces, $\Omega_\theta$ can be expanded as a
linear combination of $\Omega_m$. Finally, we have
\begin{equation}
\label{superfluidpole}
\Omega_{FL}=\sum_\theta\Omega_\theta=\Omega_{\pi_0}+\sum_m
c_m\Omega_m
\end{equation}
with the coefficients
\begin{equation}
\label{cm} c_m =\sum_\theta |\langle\theta|m\rangle|^2=
\sum_\theta{\overline{\cal M}_{mm}(M_\theta)\over
\sum_n\overline{\cal M}_{nn}(M_\theta)},
\end{equation}
where  $\overline{\cal M}$ is a matrix defined in the three
dimensional meson isospin subspace,
\begin{equation}
\label{matrix} \overline{\cal
M}(M_\theta)={\text{det}\left(1-2G\Pi(M_\theta,{\bf 0})\right)\over
1-2G\Pi(M_\theta,{\bf 0})}.
\end{equation}

\begin{figure}[!hbt]
\includegraphics[width=0.5\textwidth]{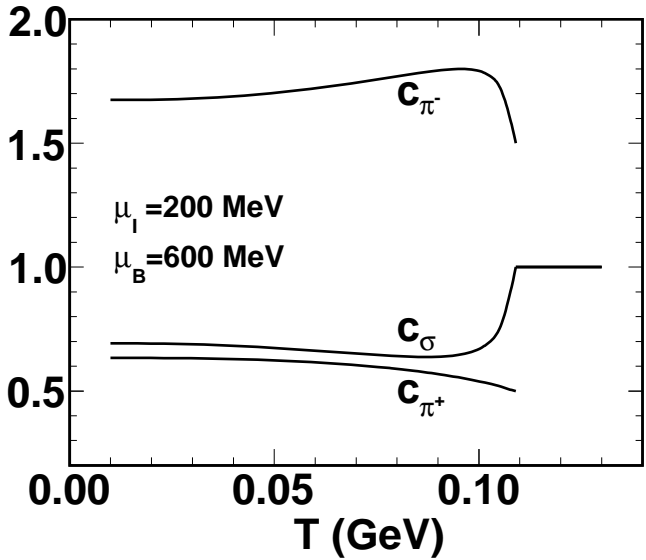}
\caption{ The linear coefficients $c_m$ for the transformation from
the eigen states of Hamiltonian to the eigen states of isospin. }
\label{fig1}
\end{figure}

It is easy to see the normalization condition for the coefficients,
$\sum_m c_m=\sum_\theta =3$, it means that only two of the three
coefficients are independent. The coefficients $c_m$ as functions of
temperature at fixed chemical potentials are shown in
Fig.\ref{fig1}. Their strong deviation from unit indicate a strong
mixing of $\sigma, \pi_+, \pi_-$ in the pion superfluid. For the
Goldstone mode $\overline \pi_+$, its linear combination is
$|\overline \pi_+\rangle=1/\sqrt
2(|\pi_+\rangle-|\pi_-\rangle)$\cite{hao}, and the two fractions are
equal and medium independent. Therefore, at the critical point the
coefficient $c_{\pi_+}$ jumps up from 0.5 to 1 and $c_{\pi_-}$ drops
down from 1.5 to 1. For $T>T_c$ in the normal phase, all the three
coefficients are unit. For $\mu_I=200$ MeV and $\mu_B=600$ MeV in
Fig.\ref{fig1}, $T_c$ is about 110 MeV. It is necessary to note that
the discontinuity of the coefficients $c_{\pi_+}$ and $c_{\pi_-}$
happens on the whole phase transition border. However, when we
approach to the border from the pion superfluid side, the pion
condensate goes to zero continuously, and this can smooth the
thermodynamics on the border, see the calculations below.

Now we use the trace anomaly relation (\ref{thermal}) to calculate
the gluon condensate, under the assumption that the NJL model can
describe reasonably well the QCD thermodynamics in the pion
superfluid. From the thermodynamic potential relative to the vacuum
$\overline\Omega(T,\mu_B,\mu_I)=\Omega(T,\mu_B,\mu_I)-\Omega(0,0,0)$,
we obtain the pressure $p=-\overline\Omega$ and energy density
$\epsilon=-p+Ts+\mu_Bn_B+\mu_In_I$ with the entropy density
$s=-\partial\overline\Omega/\partial T$, baryon number density
$n_B=-\partial\overline\Omega/\partial\mu_B$ and isospin number
density $-\partial\overline\Omega/\partial\mu_I$.

Before we make numerical calculations, we first determine the
parameters in the model. Since the NJL model is non-renormalizable,
we can employ a hard three momentum cutoff $\Lambda$ to regularize
the gap equations for quarks and pole equations for mesons. In the
following numerical calculations, we take the current quark mass
$m_0 = 5$ MeV, the coupling constant $G = 4.93$ GeV$^{-2}$ and the
cutoff $\Lambda = 653$ MeV. This group of parameters corresponds to
the pion mass $m_\pi = 134$ MeV, the pion decay constant $f_\pi= 93$
MeV and the effective quark mass $M_q = 310$ MeV in vacuum.

\begin{figure}[!hbt]
\includegraphics[width=0.5\textwidth]{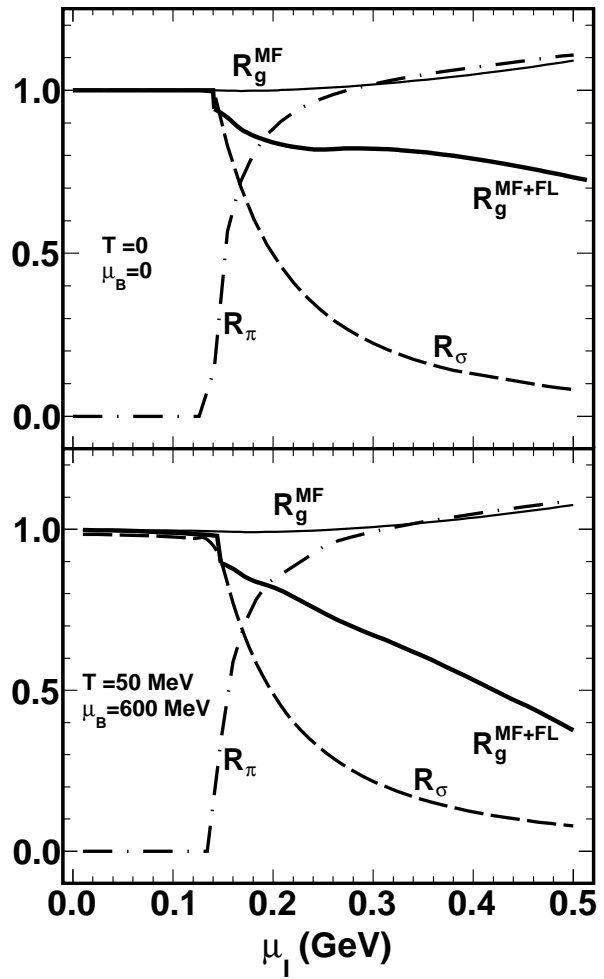}
\caption{ The scaled gluon condensates $R_g^{MF}$ in mean field
approximation and $R_g^{MF+FL}$ including quantum fluctuations. As a
comparison, we showed also the scaled chiral condensate $R_\sigma$
and pion condensate $R_\pi$ at mean field level. The top panel and
bottom panel correspond respectively to $T=\mu_B=0$ and $T=50$ MeV,
$\mu_B=600$ MeV.}
\label{fig2}
\end{figure}

We show in Fig.\ref{fig2} the ratios for gluon, chiral and pion
condensates, $R_g=\langle{\alpha_s\over
\pi}G^a_{\mu\nu}G_a^{\mu\nu}\rangle/\langle{\alpha_s\over
\pi}G^a_{\mu\nu}G_a^{\mu\nu}\rangle_0, R_\sigma=\sigma/\sigma_0$ and
$R_\pi=\pi/\sigma_0$, where $\langle{\alpha_s\over
\pi}G^a_{\mu\nu}G_a^{\mu\nu}\rangle_0$ and $\sigma_0$ are the
condensates in vacuum, and $\langle{\alpha_s\over
\pi}G^a_{\mu\nu}G_a^{\mu\nu}\rangle, \sigma$ and $\pi$ are the total
condensates including the vacuum and matter parts. To reduce the
model dependence and focus on the medium effect, we take an
empirical value for the vacuum part of the gluon condensate,
$\langle{\alpha_s\over \pi}G^a_{\mu\nu}G_a^{\mu\nu}\rangle_0$=(360
MeV)$^4$\cite{reinders} (the value of the vacuum part will not
change the trend of the gluon condensate in the medium). At
$T=\mu_B=0$ in the top panel of Fig.\ref{fig2}, the ratio
$R_g^{MF+FL}$, calculated with the total thermodynamic potential
$\Omega=\Omega_{MF}+\Omega_{FL}$, is a constant in the normal phase
with $\mu_I<m_\pi$ and drops down monotonously in the pion
superfluid phase with $\mu_I>m_\pi$. Therefore, the behavior of the
gluon condensate at finite isospin density is qualitatively the same
as in the case at finite temperature: The broken scale symmetry of
the system is gradually restored in hot and dense medium. However,
in mean field approximation, the gluon condensate behaves very
differently. The ratio $R_g^{MF}$ decreases slightly only in the
beginning of the pion superfluid and then goes up monotonously and
even exceeds the vacuum value when $\mu_I$ is high enough. For
$T=50$ MeV and $\mu_B=600$ MeV shown in the bottom panel of
Fig.\ref{fig2}, while the mean field calculation is changed
slightly, the finite temperature and baryon chemical potential
effect results in stronger meson fluctuations, and the ratio
$R_g^{MF+FL}$ drops down much faster.

The mean field result can be understood by the competition between
the chiral and pion condensates. At mean field level, the NJL
Lagrangian density can be written as
\begin{eqnarray}
\label{njl2} \mathcal {L}_{MF} &=& \overline\psi\left(i\gamma^\mu
\partial_\mu-m+\mu\gamma_0\right)+2G\left[\sigma\overline\psi\psi+\pi\overline\psi
i\gamma_5\tau_1\psi\right]\nonumber\\
&-&G(\sigma^2+\pi^2),
\end{eqnarray}
and the corresponding trace of the Noether current for the scalar
transformation is $T_\mu^\mu =
m\overline\psi\psi-2G\left(\sigma\overline\psi\psi+\pi\overline\psi
i\gamma_5\tau_1\psi\right)$. Taking the identification of the trace
of the energy-momentum tensor in QCD and in the NJL model, the gluon
condensate is characterized only by the two condensates,
\begin{eqnarray}
\label{mf2} \langle {\alpha_s\over \pi}
G^a_{\mu\nu}G_a^{\mu\nu}\rangle_{T, \mu} &=& \frac{16}{9}\
G\left(\sigma^2-\sigma^2_0+\pi^2\right)\nonumber\\
&=&{16\over 9}G\sigma_0^2\left(R_\sigma^2+R_\pi^2-1\right).
\end{eqnarray}
In the pion superfluid phase, the two ratios $R_\sigma$ and $R_\pi$
behave in an opposite way, $R_\sigma$ drops down but $R_\pi$ goes
up, and the trend of the gluon condensate is controlled by the
competition between the chiral and pion condensates. When $\mu_I$ is
above but close to the critical point $\mu_I^c=m_\pi$, the chiral
and pion condensates are equally important and their competition may
result in a possible decreasing gluon condensate. However, when
$\mu_I$ is large enough, the chiral condensate becomes small and the
pion condensate dominates the system. In this case, the gluon
condensate increases with increasing pion condensate.

It is necessary to emphasize again that the trace anomaly relation
(\ref{thermal}) between the gluon condensate and the thermodynamics
of the system is valid only at quantum level. At classical or mean
field level, the relation is not true, and the scale symmetry of QCD
is only explicitly broken by the current quark mass $m$, $\langle
T^\mu_\mu\rangle = m\langle\bar\psi\psi\rangle$. In the NJL model,
the quantum fluctuations or the meson modes can not be neglected. At
mean field level, there are only quarks in the model which control
the thermodynamics only at high temperature and density. At moderate
temperature and density around the chiral and pion superfluid phase
transitions, both quarks and mesons are important. At low
temperature and density, mesons become the dominant contribution to
the thermodynamics. Therefore, we need quantum fluctuations to
describe the system in the whole temperature and density region.

In summary, we have studied the gluon condensate beyond mean field
approximation in a pion superfluid described by the NJL model. Since
the trace anomaly relation is valid only at quantum level, the
quantum fluctuations in the model must be considered in the
calculation of gluon condensate. At classical or mean field level,
the growing pion condensate in the superfluid leads to a surprising
increase of the gluon condensate. However, when the quantum
fluctuations are included, the meson contribution dominates the
thermodynamics of the system at low and intermediate temperature and
density, and the gluon condensate becomes to decrease gradually in
the pion superfluid. Therefore, the scale symmetry can be restored
at both finite temperature and density.

\appendix {\bf Acknowledgement:} The work is supported by the NSFC
grant Nos. 10735040, 10847001, 10975084 and 11079024.

\end{document}